\documentclass{PoS}

\usepackage{amsmath}

\newcommand{\e}{\mathrm{e}}

\newcommand{\rO}{\mathrm{O}}
\newcommand{\<}{\langle}
\newcommand{\z}{\rangle}

\newcommand{\ev}[1]{\bigl\<#1\bigr\z}

\newcommand{\plotwidth}{.45\textwidth}
\newcommand{\onecol}[2]{
        \begin{minipage}[t]{#1}{#2\vfill} \end{minipage}
        }

\title{Scale $r_0$ and the static potential from the CLS lattices}

\ShortTitle{Scale $r_0$ and the static potential from the CLS lattices}

\author{
   \includegraphics[width=0.25\linewidth]{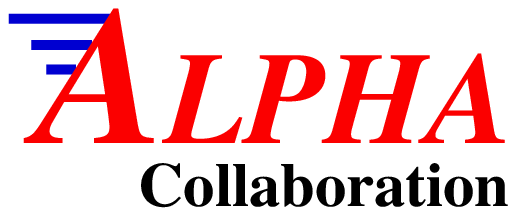}
   \hfill
   \onecol{3cm}{\vspace{-4.5em}\it
      BUW-SC 2010/4\\
      WUB/10-37
   }
   \vspace{1cm}
}
\author{\speaker{Bj\"orn Leder}$^{a,b}$,
        Francesco Knechtli$^a$\\
        \llap{$^a$}Department of Physics, Bergische Universit\"at Wuppertal\\
                   Gaussstr. 20, D-42119 Wuppertal, Germany\\
        \llap{$^b$}Department of Mathematics, Bergische Universit\"at Wuppertal\\
                   Gaussstr. 20, D-42119 Wuppertal, Germany\\
        E-mail: \email{leder@physik.uni-wuppertal.de}
}

\abstract{We report on the measurement of the static potential and the scale $r_0$ from
          HYP-smeared Wilson loops in two flavour QCD. We analyse the quark mass dependence
          of the potential and $r_0$ at three lattice spacings. We also compare the QCD static
          potential around distance $r_0$ with the static potential obtained from potential
          models.}

\FullConference{The XXVIII International Symposium on Lattice Field Theory, Lattice2010\\
		June 14-19, 2010\\
		Villasimius, Italy}

\begin{document}

\section{Introduction}
\vspace{-2ex}

The potential between two static colour charges in QCD is a quantity rich in features.
At large distances $r$ between the two charges the phenomenon of string breaking
is observed for the theory with dynamical quarks \cite{Bali:2005fu}. The
quenched theory is expected to asymptotically coincide with bosonic effective string
theory for $r\to\infty$ \cite{Luscher:2002qv}. At intermediate distances a scale
$r_0$ \cite{Sommer:1993ce} can be defined which is readily computed
in numerical simulations and can be related to phenomenological models of
quarkonium. This scale can then be used to set the overall scale of a simulation
and/or set the relative scale for simulations at different lattice spacings. Finally, at
short distances the shape of the potential is described by renormalised perturbation theory.
In fact, by taking the second derivative with respect to the distance a renormalised coupling
can be defined. In this way the static potential connects
the non-perturbative regime and the perturbative regime and is also a quantity where
one expects large effects of dynamical quarks, i.e., large differences between
the quenched and the unquenched result.

Here we report on an ongoing effort to measure the static potential on the
configuration ensembles generated by CLS 
(Coordinated Lattice Simulations)\footnote{https://twiki.cern.ch/twiki/bin/view/CLS/WebHome}.
They were generated with the DD-HMC software package\footnote{http://luscher.web.cern.ch/luscher/DD-HMC},
which implements 
two degenerated flavours of improved Wilson fermions and the Wilson gauge action.
There are ensembles at three different values of $\beta$ (i.e., three lattice spacings)
and several values of the sea quark mass. The first objective is to provide the scale $r_0$.
It can be compared to other methods of scale determination \cite{Brandt:2010ed}, used to compare dimensionless
quantities among different collaborations, perform scaling analysis and preliminarily set
the overall scale in physical units\footnote{Because of the uncertainty coming from
the phenomenological models used to give $r_0$ in physical units, this is somewhat
unsatisfying and should be seen as an intermediate step.}. Since the data presented
here are not based on all the available statistics and some details of the analysis
might change, the results presented here are preliminary.

The report is organised as follows. In the next section the techniques used to extract
the static potential with low statistical fluctuations and small systematic errors
are summarised (see \cite{Donnellan:2010} for more details). In section \ref{scale} the
scale $r_0$ is determined and extrapolated to vanishing sea quark mass.
In section \ref{physics} applications beyond scale setting are collected, i.e.,
a comparison of the quark mass dependence of $r_0$ with other collaborations and the
determination of a renormalised coupling.

\section{The static potential}
\label{techniques}
\vspace{-2ex}

To determine the static potential $V(r)$ we measure rectangular Wilson loops. They have
extension $t$ and $r$ in temporal and spatial direction respectively. Schematically,
we measure
\begin{equation}
   C(t) = \ev{{\scriptstyle t}\, \overset{r}{\underset{r}{\Box}}\, {\scriptstyle t}}\,,\quad r\; \text{fixed}\,,
   \label{wilson_loop}
\end{equation} 
where the brackets denote the expectation value with respect to the two flavour QCD
measure and the square represents the product of link variables around a closed 
rectangular path. This is equivalent to a static quark--anti-quark pair at spatial
separation $r$ propagating distance $t$ in time. For large euclidean time separations
the signal is dominated by the ground state, which coincides with the static potential
\begin{equation}
   C(t) \overset{t\to\infty}{=} A(r)\;\e^{-V(r)\,t}\,,\quad r\; \text{fixed}\,.
   \label{large_time}
\end{equation}
A straightforward determination of $V(r)$ from this definition with the original
link variables of the configurations is bound to suffer from large statistical and
systematic errors for two reasons. First, the signal to noise ratio will decrease
exponentially for large $r$ and $t$ due to ultraviolet fluctuations. Second, the overlap
with the ground state (encoded in $A(r)$ in eq. \eqref{large_time}) will be poor.

To remove the ultraviolet fluctuations the original link variables are replaced by
smeared links. To this end one level of hypercubic smearing \cite{Hasenfratz:2001hp} with parameters 
$\alpha_1=1.0$, $\alpha_2=1.0$, $\alpha_3=0.5$, referred to as HYP2 \cite{DellaMorte:2003mn} was found
to give the best result. The whole analyses was carried out with a second parameter set
$\alpha_1=0.75$, $\alpha_2=0.6$, $\alpha_3=0.3$, referred to as HYP1. When not specified,
the parameter set HYP2 is used.
The smearing of the temporal links can be understood in terms of choosing an
action for the static quarks and the smearing of the spatial links in terms of a
redefinition of the operator that creates the static
quark--anti-quark pair (see \cite{Donnellan:2010} for a detailed derivation).
Using smeared links greatly improves the signal to noise ratio at large time separations.

\begin{figure}[tb]
\begin{minipage}[b]{\plotwidth}
   \centering
   \includegraphics[width=\linewidth]{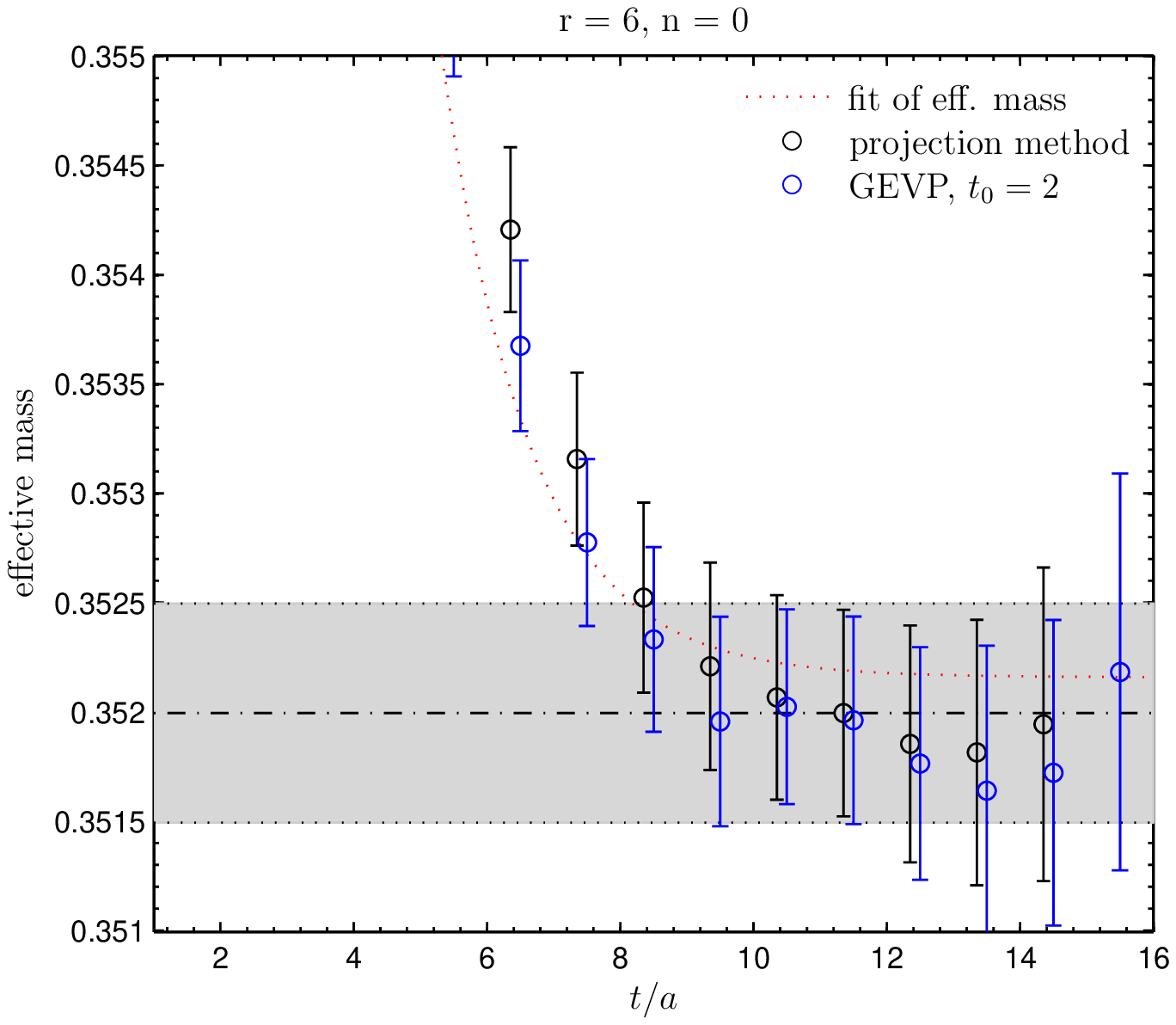}
   \caption{Effective ``mass'' plot for the Wilson loop correlator matrix for
            $r=6a$ and $t_0=2a$. See text for a detailed explanation.}
   \label{fig_eff_mass}
\end{minipage}  
\hfill
\begin{minipage}[b]{\plotwidth}
   \centering
   \includegraphics[width=0.94\linewidth]{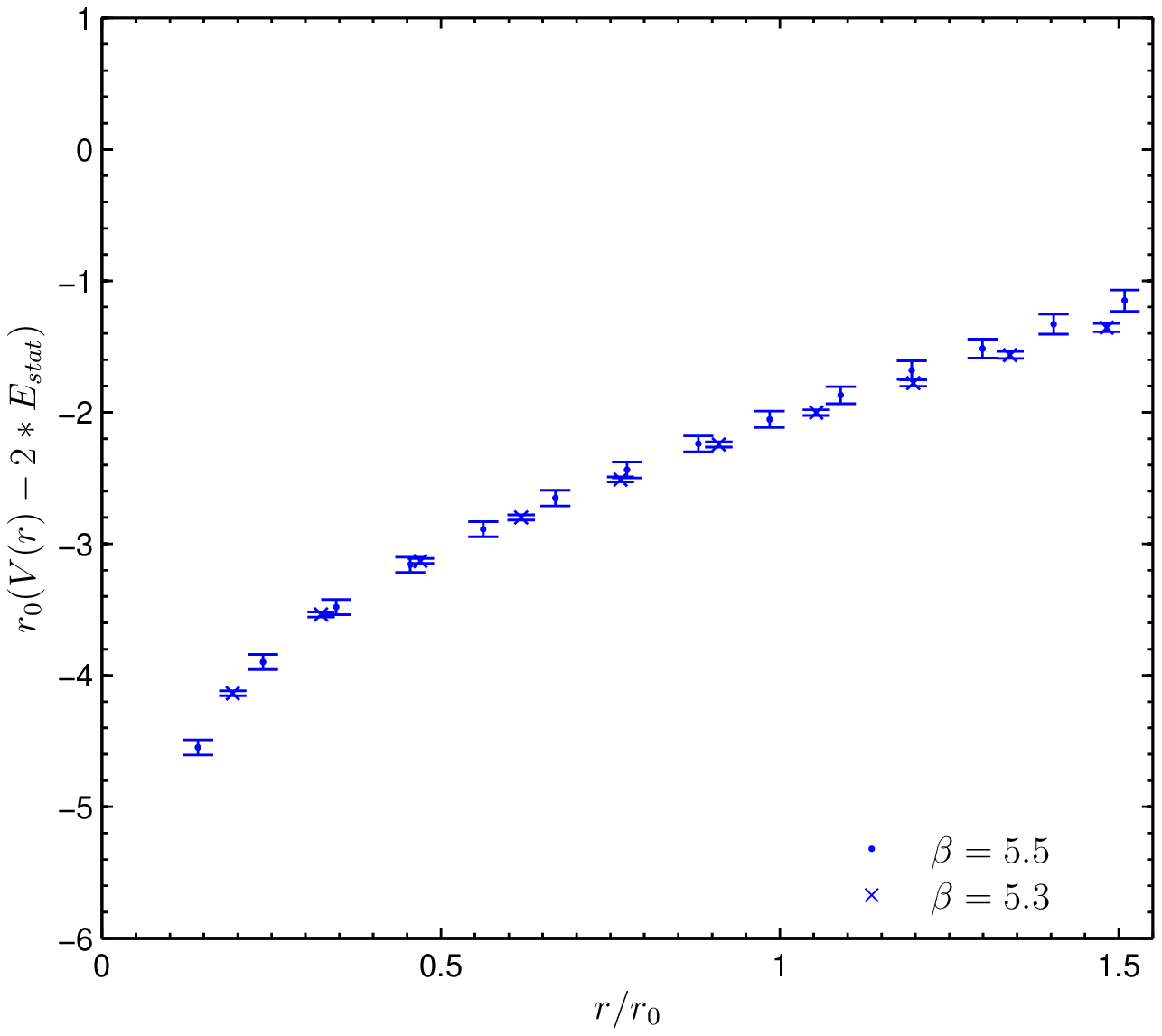}
   \caption{Static potential at two values of the lattice spacing. The static energy
            $E_{\text{stat}}$ \cite{lat10:hqet} is subtracted to obtain a renormalised quantity.}
   \label{fig_potential}
\end{minipage}  
\end{figure}

Since the exact wave function of the desired ground state is unknown the usage of
a variational method is mandatory to improve the overlap. 
As already explained further smearing the spatial links on the left hand side of eq.
\eqref{wilson_loop} is equivalent to a redefinition of the operator that creates
the static quark--anti-quark pair. Intuitively, by smearing the generated sate becomes
more and more extended. This way one obtains a correlator matrix
\begin{equation}
   C_{lm}(t) = \ev{{\scriptstyle t}\, \overset{r,m}{\underset{r,l}{\Box}}\, {\scriptstyle t}}\,,\quad r\; \text{fixed}\,.
   \label{wilson_loop_matrix}
\end{equation} 
In particular the indices $l,m=1,\dots,M$ specify the $n_{l,m}$ levels of spatial HYP
smearing with parameters $\alpha_2=0.6$, $\alpha_3=0.3$\footnote{Only spatial links
are involved, thus only two parameters are needed.} that are applied to the spatial
links. Throughout this report we use $M=4$ and $n_l=l-1$.

From the correlator matrices \eqref{wilson_loop_matrix} effective ``masses'' are
extracted with two different methods. The starting point in both cases is the
generalised eigenvalue problem (GEVP)
\begin{equation}
   C(t)\, \psi_\alpha(t,t_0) = \lambda_\alpha(t,t_0)\, C(t_0)\, \psi_\alpha(t,t_0)\,, \quad \alpha=0,\dots,M-1\,.
   \label{gevp}
\end{equation}
From the generalised eigenvalues one directly obtains
\begin{equation}
   E_\alpha(t+\tfrac{a}{2},t_0) \equiv \ln(\lambda_\alpha(t,t_0)/\lambda_\alpha(t+a,t_0))
                                 = E_\alpha + \beta_\alpha\e^{-(E_M-E_\alpha)(t+\tfrac{a}{2})} + \dots\,,
   \label{gev}
\end{equation}
where $E_0\equiv V(r)$ and the higher states contributions are expected to die out
with large $E_M-E_\alpha$ or faster. In \cite{Blossier:2009kd} the authors were able
to prove this for $t\leq 2 t_0$.
In fig. \ref{fig_eff_mass} $E_0(t+\tfrac{a}{2},t_0)$ is plotted for an
intermediate $r$ and $t_0=2a$ (blue circles)\footnote{Statistical errors are determined
using the method and program of \cite{Wolff:2003sm}.}. In order to quantify the contribution
of higher states we perform a fit with the two terms of the right hand side of
eq. \eqref{gev} including data points for $\alpha=0,1$ and $t\leq 2 t_0$
(red dotted curve). The term modelling the higher states is used to estimate
the systematic error due to them when extracting the ground state at a given time $t$
(see below).

Finally, the projection method of \cite{Guagnelli:1998ud} is used to obtain a
second determination of $E_\alpha(t+\tfrac{a}{2},t_0)$. There the
correlator matrices are projected to the ground state generalised eigenvector
\mbox{$v=\psi_0(t_0+a,t_0)$}
\begin{equation}
   f(t) = v^T\, C(t)\, v\,.
   \label{projection}
\end{equation}
The resulting numbers are locally fitted by a single exponential, i.e. three successive
values $f(t_p)$, $t_p=t-a,t,t+a$ are fitted to $f(t_p)=b\,\e^{-E_0(t,t_0)t_p}$
(black circles in fig. \ref{fig_eff_mass}).

\begin{figure}[t]
\begin{minipage}[b]{\plotwidth}
   \centering
   \includegraphics[width=0.97\linewidth]{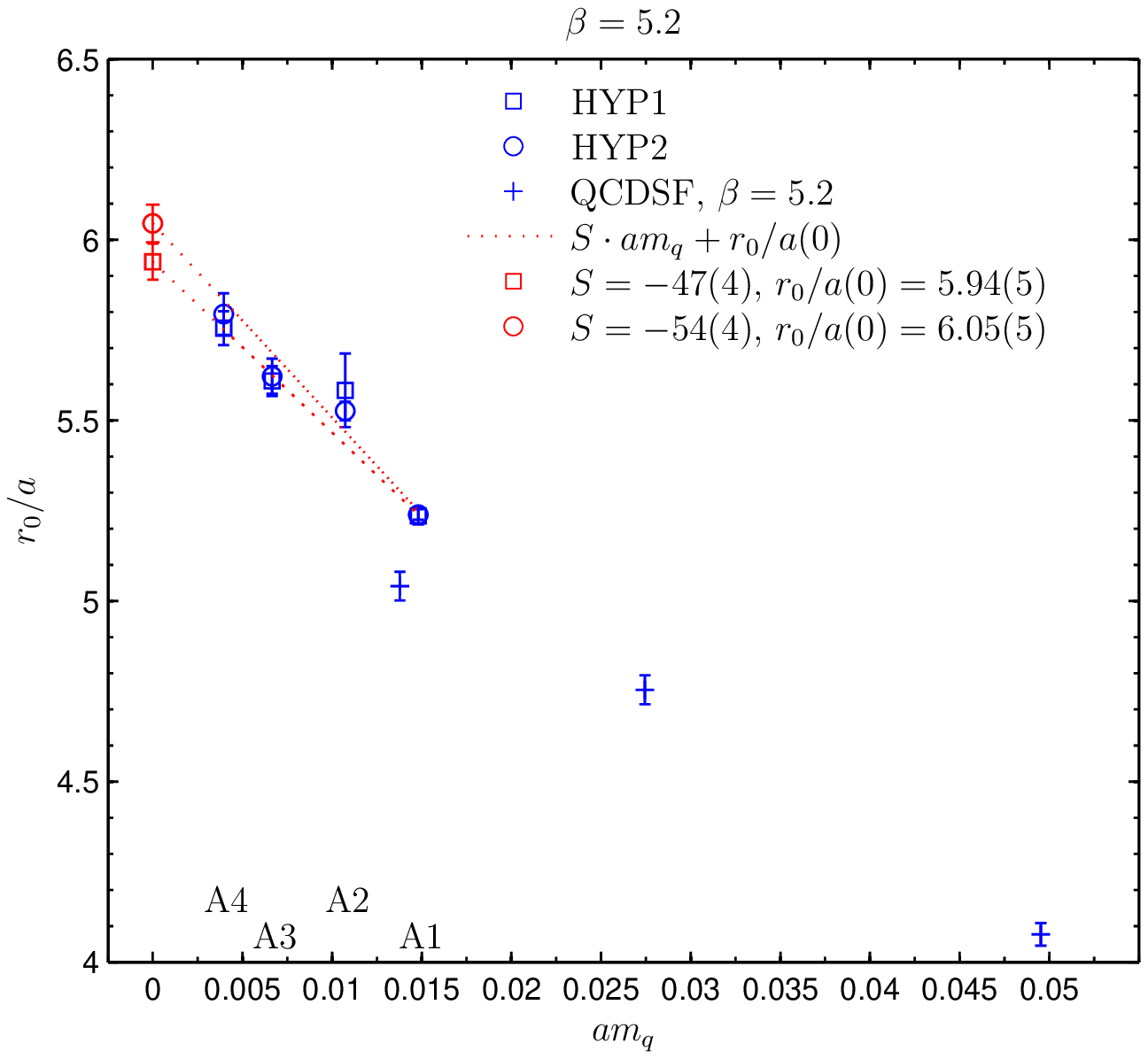}
   \caption{Chiral extrapolation of $r_0/a$ at \mbox{$\beta=5.2$}. QCDSF data
            from \cite{Brommel:2006ww} for comparison.}
   \label{fig_r0_kappa1}
\end{minipage}  
\hfill
\begin{minipage}[b]{\plotwidth}
   \includegraphics[width=\linewidth]{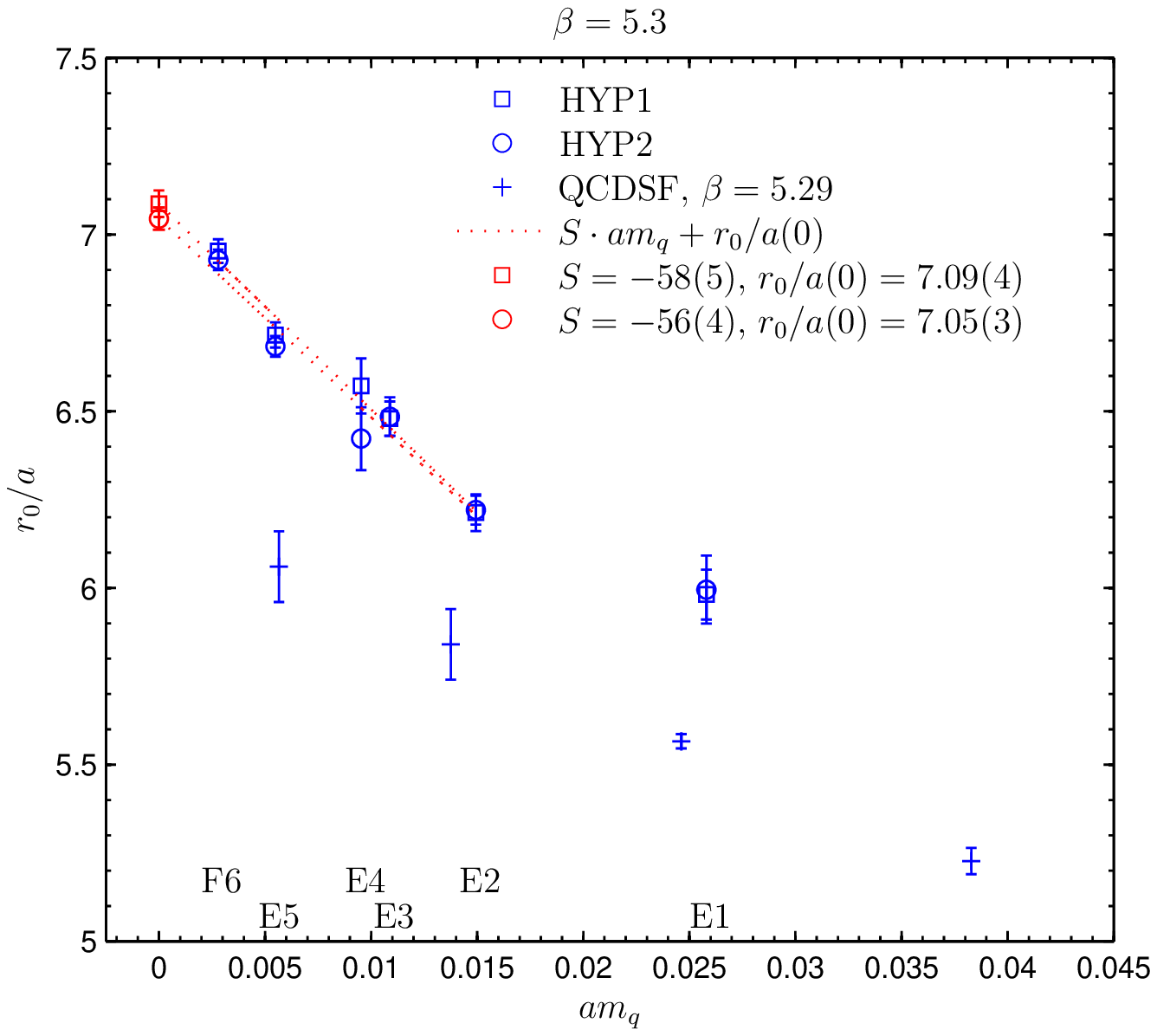}
   \caption{Chiral extrapolation of $r_0/a$ at \mbox{$\beta=5.3$}. QCDSF data
            from \cite{Brommel:2006ww} for comparison.}
   \label{fig_r0_kappa2}
\end{minipage}  
\end{figure}

With the parameters as in this report we found the projection method to be slightly
more stable, i.e. to exhibit longer plateaus. Therefore we extract the static potential
from the black circle in fig. \ref{fig_eff_mass} with the smallest sum of statistical
and systematic error (black dashed-dotted curve and grey error band).

Applying this procedure to all values of $r$ we obtain the static potential as a
function of $r$. In fig. \ref{fig_potential} we plot the result (made dimensionless
by multiplying with $r_0$, see below) for two values of
the lattice spacing at roughly the same sea quarks mass. The plot shows that
lattice artefacts are small.

\section{Scale $r_0$}
\label{scale}
\vspace{-2ex}

The scale $r_0$, introduced in \cite{Sommer:1993ce}, is defined in terms of the static
force $F(r)=V'(r)$ by solving
\begin{equation}
   r^2\, F(r)\big|_{r=r_0}=1.65\,.
   \label{def_r0}
\end{equation}
Its physical value is $r_0\approx 0.5\; \text{fm}$, thus it is sensitive to the non-perturbative
character of the theory. Off the lattice it can only be determined through phenomenological
potential models. On the lattice the static force is computed from the potential
as the finite difference
\begin{equation}
   F(r_I) = \tfrac{1}{a}[V(r)-V(r-a)]\,,
   \label{def_Flat}
\end{equation}
where $r_I=r-a/2+\rO(a^2)$ is chosen such that in \eqref{def_Flat} at tree level
all lattice artefacts cancel out. To solve \eqref{def_r0} the force is then 
locally parametrised by $a^2\,F(r)=f_0+f_2\,a^2/r^2$. Taking the two values that enclose the
solution, the parametrisation is uniquely determined as well as $r_0/a$. This
lattice definition of $r_0/a$ ensures that it is a smooth function of the bare
parameters $\beta$ and $m_0$. To estimate the systematic uncertainties due to the
parametrisation the same procedure is repeated with an additional term $f_4\, a^4/r^4$
and three successive $r$-values. They are found to be negligible.

\begin{figure}[tb]
\begin{minipage}[b]{\plotwidth}
   \centering
   \includegraphics[width=\linewidth]{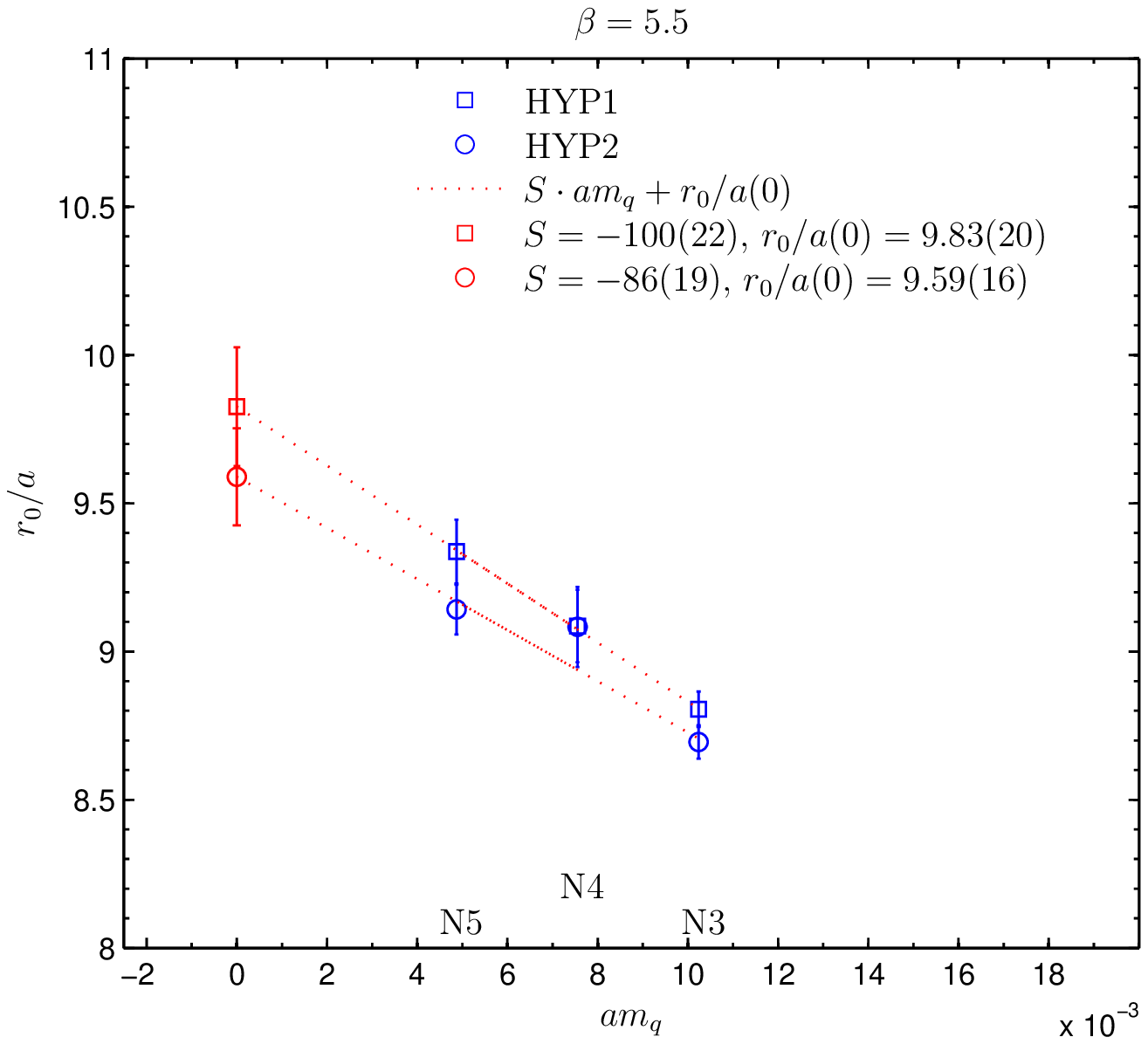}
   \caption{Chiral extrapolation of $r_0/a$ at \mbox{$\beta=5.5$}.}
   \label{fig_r0_kappa3}
\end{minipage}  
\hfill
\begin{minipage}[b]{\plotwidth}
   \centering
   \includegraphics[width=\linewidth]{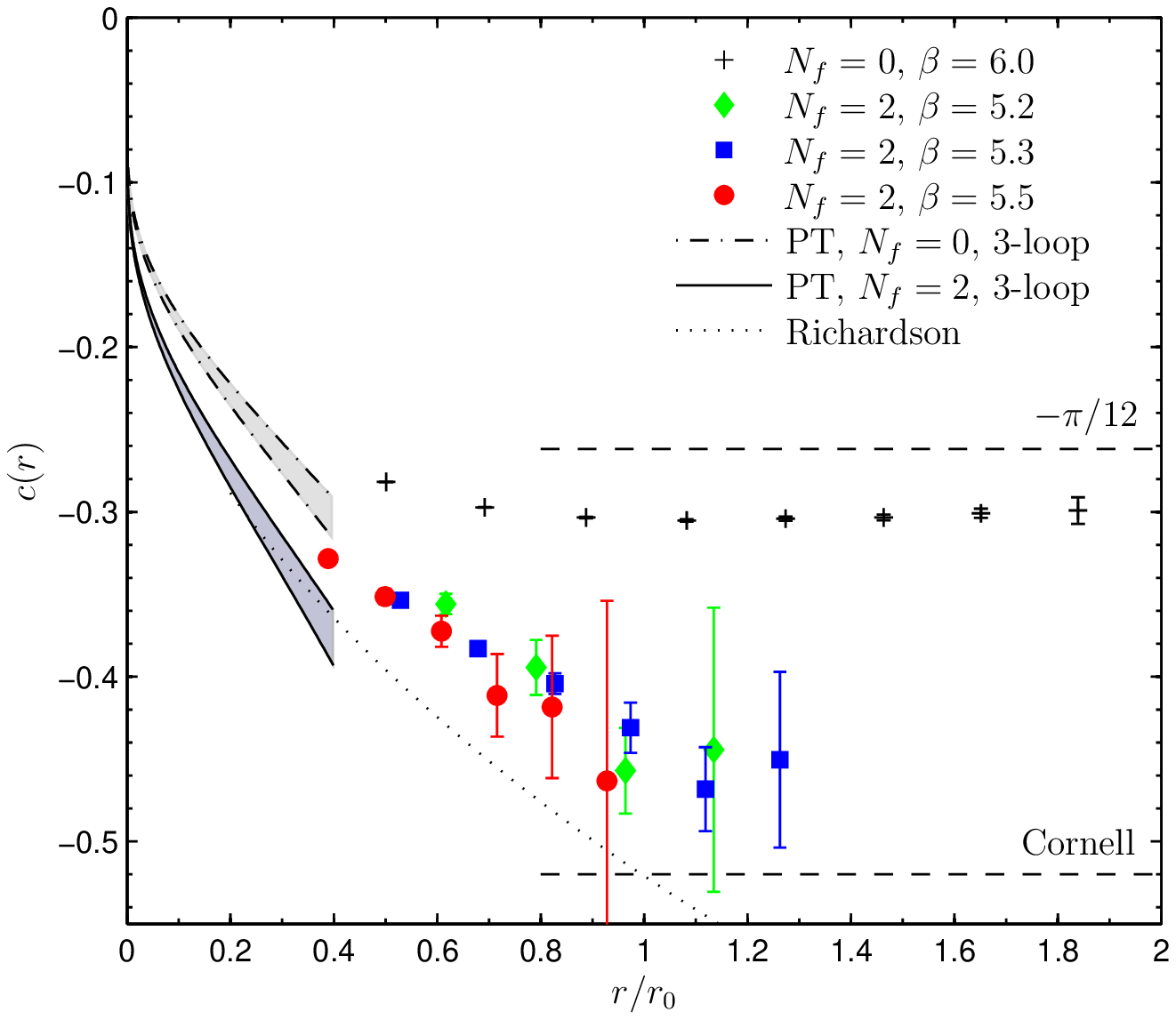}
   \caption{Renormalised quantity $c(r)$. $N_f=2$ data are at $r_0 M_{PS}\approx 1$.
            See text for details.}
   \label{fig_c}
\end{minipage}  
\end{figure}

We analysed three run sets at three different lattice spacings with
$\beta=5.2\,, 5.3\,, 5.5$. At each lattice spacing there are runs at $3$--$6$ 
quark masses.  For a scale setting we decided to extrapolate to the chiral point.
In figs. \ref{fig_r0_kappa1}--\ref{fig_r0_kappa3} the measured values of $r_0/a$
are plotted versus the subtracted quark mass $am_q=am_0-am_{\text{cr}}$. Here 
we give the result for both parameter sets HYP1 and HYP2. The
extrapolation to $am_q=0$ is done via a linear fit. The result of the fits
are given in the legends of the plots. The HYP2 results are compatible with the HYP1
ones and have the smaller error. Thus we list here for $r_0/a(\beta)$ in the chiral limit:
\begin{equation}
   r_0/a(5.2) = 6.05(5)\,,\quad r_0/a(5.3) = 7.05(3)\,,\quad r_0/a(5.5) = 9.59(16)\,.
   \label{result_r0}
\end{equation}
Using our new determination of $r_0/a$ we get a preliminary update on the $\Lambda$-parameter
of \cite{DellaMorte:2003mn}: 
\begin{equation}
   r_0 \Lambda_{\overline{\text{MS}}}^{Nf=2} = 0.73(3)(5)\,, 
\end{equation}
where the first error comes from $r_0/a$ and the second from the running of the coupling.

In figs. \ref{fig_r0_kappa1} and \ref{fig_r0_kappa2} we also plot data from the
QCDSF collaboration \cite{Brommel:2006ww}. Since they are using exactly the same action, the results
at coinciding bare parameters should agree within errors. The most tension is observed
between their lightest point and the heaviest of this report at $\beta=5.2$ in 
fig. \ref{fig_r0_kappa1}. Also in fig. \ref{fig_r0_kappa2} the difference is hardly
explainable by the difference in the $\beta$-values ($5.3$ versus $5.29$). The most
plausible explanation is that they used a global fit including many terms to model
the potential (and thus the force)\footnote{private communication with P. Rakow}
over a wide range of $r$ values. This way the
statistical error in $r_0/a$ is reduced, but at the price of large systematic
uncertainties.

\section{Physics results}
\label{physics}
\vspace{-2ex}

\begin{figure}
   \centering
   \includegraphics[width=\plotwidth]{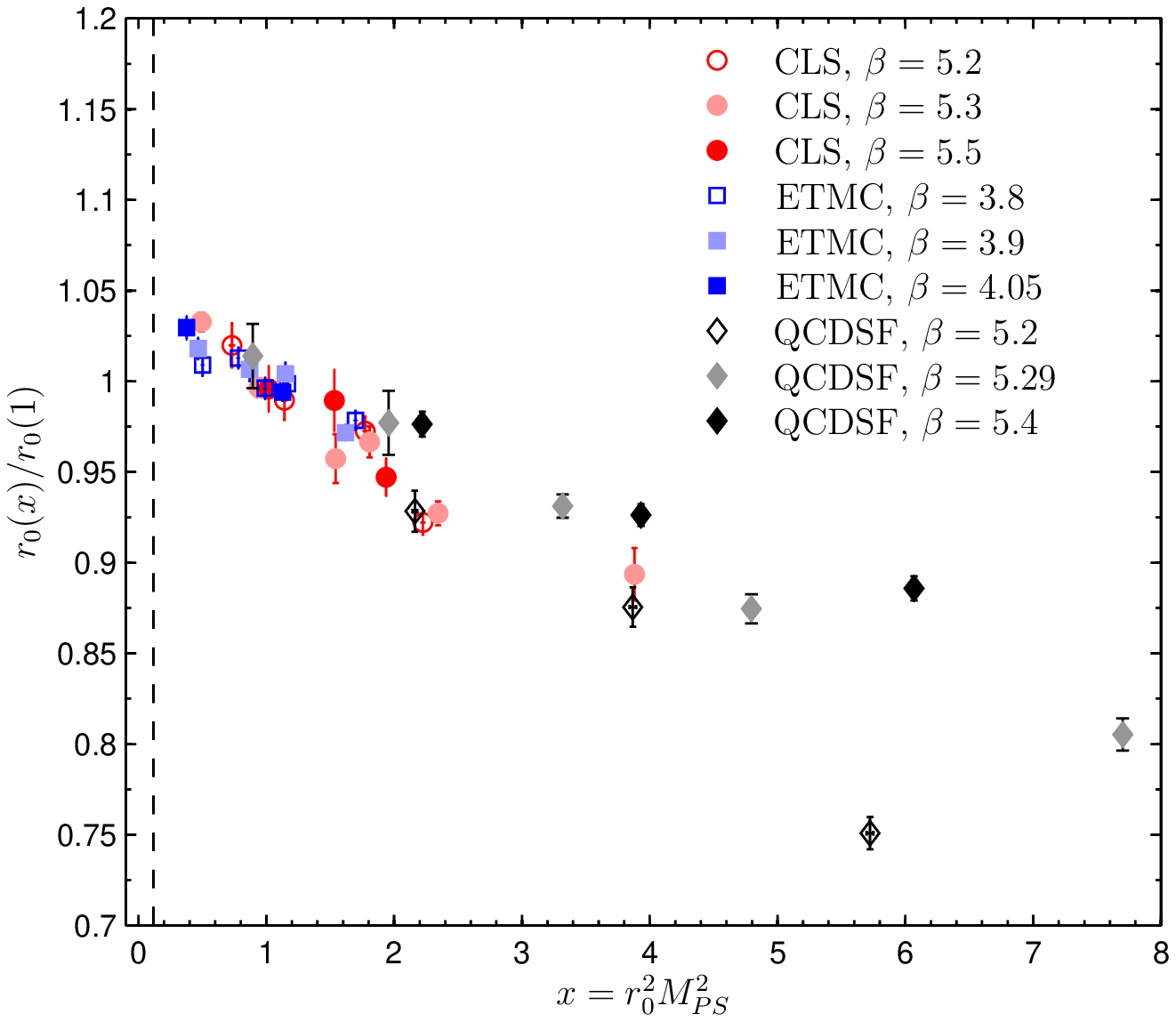}\hfill
   \includegraphics[width=\plotwidth]{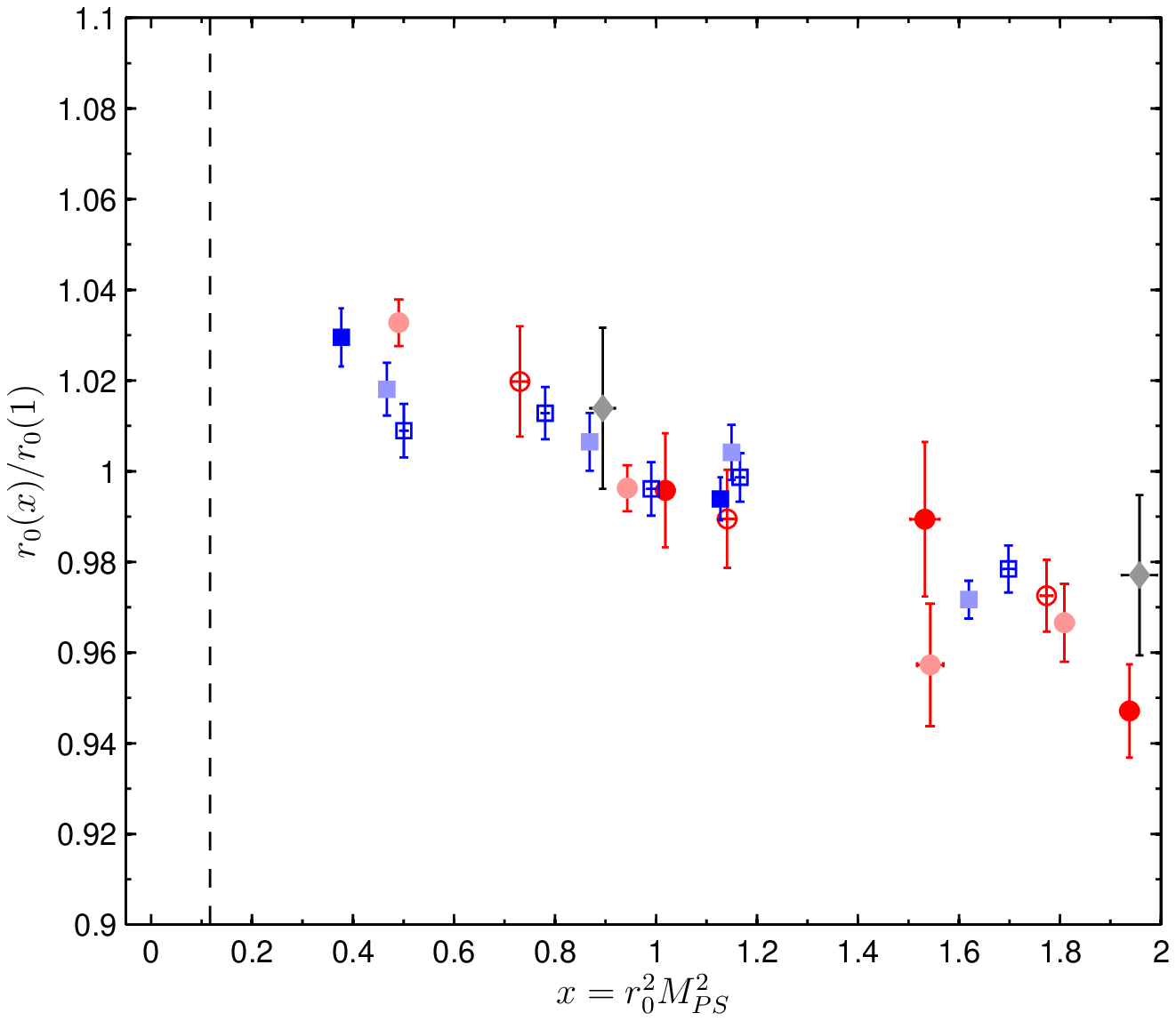}
   \caption{Plot of dimensionless and physical quantities comparing different efforts
            to simulate two flavour QCD. The right panel is a blow-up of the region
             $x=0\dots 2$.
            \vspace{-2ex}}
   \label{fig_r0_mps}
\end{figure}

First, we present a quantity for which large effects of dynamical quarks can be observed.
A renormalised, physical quantity can be defined in terms of the derivative
of the static force
\begin{equation}
   c(r)=\tfrac{1}{2}r^3F'(r)\,.
   \label{def_c}
\end{equation}
In \cite{Luscher:2002qv} it was determined with high precision in pure gauge
theory. With the data presented here we are able to compare the pure gauge case 
to the $N_f=2$ theory. On the lattice we write $c$ in terms of a finite difference
\begin{equation}
   c(\tilde{r}) = \tfrac{1}{2a^2}\tilde{r}^3[V(r+a)+V(r-a)-2V(r)]\,,
   \label{def_clat}
\end{equation}
where $\tilde{r}=r+\rO(a^2)$ is chosen such that at tree level all lattice artefacts
cancel out. In fig. \ref{fig_c} we plot $c$ for the three values of $\beta$ at
roughly the same sea quark mass ($r_0M_{PS}\approx 1$). For comparison we also plot
the $N_f=0$ data \cite{Luscher:2002qv}. The analytic curves in the plot are the
3-loop perturbative curves (dashed-dotted for $N_f=0$ and solid line for $N_f=2$,
spread due to uncertainty in the $\Lambda$-parameter)\footnote{For details
on the perturbative expressions for $c(r)$ and how it is related to a renormalised coupling
we refer the reader to \cite{Donnellan:2010}.},
the universal value $-\pi/12$
from bosonic string theory \cite{Luscher:1980fr,Luscher:1980ac} (that the $N_f=0$ data is approaching asymptotically for
$r\to\infty$), the value of $c$ in the Cornell \cite{Eichten:1979ms} potential and the curve derived from
the Richardson \cite{Richardson:1978bt} potential.

Second, we present in fig. \ref{fig_r0_mps} a comparison to other efforts simulating 
QCD with two light quarks. For
the x-axis we define a dimensionless quantity $x=r_0^2\,M_{PS}^2$ (where $M_{PS}$
is the pseudo-scalar mass). For the y-axis we define the ratio of $r_0(x)/r_0(1)$,
where $x=1$ serves as a reference point to cancel the unknown overall scale. We have
chosen to include QCDSF \cite{Brommel:2006ww} and ETMC \cite{Baron:2009wt},
because of the readily available data for $r_0$ and $M_{PS}$. The left panel shows the whole
range of data points, whereas the right panel is a blow-up of the region close
to the physical point. Assuming a physical value $r_0=0.5 \; \text{fm}$ the physical
value of $x=r_0^2\,M_{PS}^2$ is indicated by the vertical dashed line. As one can see
the spread of the points is narrowing towards the physical point,
where they start to agree within errors.

\section*{Acknowledgements}
\label{ack}
We thank Rainer Sommer for helpful discussions on various aspects of this work,
Nazario Tantalo for extensive checks of the Wilson loop measurements, Stefan Sch\"afer
for help in checking the HYP smearing and providing some $am_{\text{cr}}$ values, and Nikos
Ingres for discussions on the quantity $c(r)$. We further thank the Forschungszentrum
J\"ulich and the Zuse Institut Berlin for allocating computing resources to this project.
Part of the Wilson loop measurements were performed on the PAX cluster at DESY, Zeuthen.

\bibliography{lat10biblio}           
\bibliographystyle{h-elsevier}   

\end{document}